\documentclass[prl,twocolumn,showpacs]{revtex4-1}
\usepackage{graphicx}
\usepackage{times}
\usepackage{bm}
\usepackage{amsmath, verbatim}
\usepackage{amsfonts}

\def\lsim{\mathrel{\rlap{\lower4pt\hbox{\hskip1pt$\sim$}}
    \raise1pt\hbox{$<$}}}                
\def\gsim{\mathrel{\rlap{\lower4pt\hbox{\hskip1pt$\sim$}}
    \raise1pt\hbox{$>$}}}                

\def\be{\begin{equation}}
\def\ee{\end{equation}}
\def\bea{\begin{eqnarray}}
\def\eea{\end{eqnarray}}
\def\bse{\begin{subequations}}
\def\ese{\end{subequations}}

\def\be{\begin{eqnarray}}
\def\ee{\end{eqnarray}}

\begin{document}

\title{Topological Properties of Time Reversal Symmetric Kitaev Chain and Applications to Organic Superconductors}

\author{E. Dumitrescu and Sumanta Tewari}

\affiliation{Department of Physics and Astronomy, Clemson University, Clemson, SC 29634, USA}

\begin{abstract}
We show that the pair of Majorana modes at each end of a 1D spin triplet superconductor with total Cooper pair spin $S_x=0$ (i.e.,  $\Delta_{\uparrow\uparrow}=-\Delta_{\downarrow\downarrow}=p\Delta_0$; two uncoupled time reversed copies of the Kitaev $p$-wave chain) are topologically robust to perturbations such as mixing by the $S_z=0$ component of the order parameter ($\Delta_{\uparrow\downarrow}=\Delta_{\downarrow\uparrow}$),
transverse hopping (in quasi-1D systems), non-magnetic disorder, and also, most importantly, to time reversal breaking perturbations such as applied Zeeman fields/magnetic impurities and the mixing by the $S_y=0$ component of the triplet order parameter ($\Delta_{\uparrow\uparrow}=\Delta_{\downarrow\downarrow}$). We show that the robustness to time reversal breaking results from a hidden chiral symmetry which places the system in the BDI topological class with an integer $\mathbb{Z}$ invariant.
 Our work has important implications for the quasi-1D organic superconductors (TMTSF)$_{\rm {2}}$X (X=PF$_6$, CIO$_4$) (Bechgaard salts) which have been proposed as triplet superconductors with equal spin pairing ($\Delta_{\uparrow\uparrow},\Delta_{\downarrow\downarrow} \neq 0, \Delta_{\uparrow\downarrow}=0$) in applied magnetic fields.
\end{abstract}

\maketitle

It was shown by Read and Green \cite{Read_Green_2000} that 2D spinless $p$-wave superconductors
host zero energy Majorana fermion (MF) excitations (with second quantized operator $\gamma$ satisfying $\gamma^{\dagger}=\gamma$) in order parameter
defects such as vortices and edges. Kitaev showed \cite{Kitaev_2001} that the 1D version of the same system
(henceforth called ``Kitaev $p$-wave chain") can host MFs at the chain ends which can be
used for topological quantum computation (TQC) \cite{Nayak_2008}.
   Recently, MFs have been proposed to exist in systems closely analogous
to the spinless $p$-wave superconductors/superfluids such as heterostructures of topological insulators and $s$-wave superconductors \cite{Fu_2008}, cold fermion systems with Rashba spin orbit coupling, Zeeman field, and an attractive $s$-wave interaction \cite{Zhang_2008,Sato_2009}, and also, heterostructures of  spin-orbit coupled semiconductor thin films \cite{Sau,Long-PRB} or nanowires \cite{Long-PRB,Roman,Oreg} proximity coupled with $s$-wave superconductors and a Zeeman field. There have been recent claims of experimental observation of MFs in the semiconductor heterostructure which have
attracted considerable attention \cite{Mourik_2012,Rokhinson_2012,Deng_2012,Das_2012,Churchill_2013,Finck_2013,Alicea_2012, Leijnse_2012,Beenakker_2013,Stanescu_2013}.
In concurrent developments recent work \cite{Schnyder_2008,Kitaev_2009,Ryu_2010} has established that the quadratic Hamiltonians describing gapped topological insulators and superconductors can be classified into 10 distinct topological symmetry classes that can be characterized by certain topological invariants.
The symmetry classification of a given system is important as it provides an understanding of the effects of various perturbations on the stability of the protected surface modes such as the MFs. Generally speaking, if a given perturbation breaks a symmetry, the surface zero energy modes of the corresponding symmetry protected topological state acquire non zero energy and are removed by that perturbation.

 Although the 1D spinless Kitaev chain has un unphysical Hamiltonian, quasi-1D spin-triplet superconductivity has been proposed in a class of organic superconductors (TMTSF)$_{\rm {2}}$X (Bechgaard salts, X=PF$_6$, ClO$_4$) in the presence of applied magnetic fields \cite{Lee_2001,Lee_2003,Shinagawa_2007,Lebed_2000} (It has also been proposed recently in
Li$_{0.9}$Mo$_6$O$_{17}$ \cite{Lebed_2013}). The Bechagaard salts are quasi-one-dimensional
charge transfer salts exhibiting pressure induced superconductivity with abnormally high upper critical fields $H_{c2}$ \cite{Lee_2003}.
The precise form of the order parameter is not completely known, but there is evidence, at least in the presence of a magnetic field, that the
superconducting state is consistent with an equal-spin-pairing (ESP) $p$-wave phase \cite{Lee_2001,Lee_2003,Shinagawa_2007,Lebed_2000,Ardavan_2012,Jerome_2008,Thomale_2013}.
Such a phase with $\Delta_{\uparrow\uparrow}, \Delta_{\downarrow\downarrow} \neq 0, \Delta_{\uparrow\downarrow}=0$ realizes two independent copies of the Kitaev $p$-wave chain, one for each spin sector. By continuity from Kitaev's argument \cite{Kitaev_2001}, since the two spin sectors are uncoupled, one expects \textit{a pair} of MFs (one from each spin sector) at \textit{each end} of the chains. (The average spin-polarizations of the MFs are zero, however, since they are an equal superposition of particles and holes within a single spin sector.) Nevertheless, since for $\Delta_{\uparrow\uparrow}=-\Delta_{\downarrow\downarrow}=p\Delta_0$ (henceforth called the ``TR-symmetric Kitaev chain") the Hamiltonian is symmetric under time reversal, it may appear that the pair of MFs at a given end are protected by the TR symmetry, the topological class being DIII with a $\mathbb{Z}_2$ invariant. A consequence of this would be the MFs in (TMTSF)$_{\rm {2}}$X (or in Li$_{0.9}$Mo$_6$O$_{17}$) would acquire a gap in the presence of Zeeman fields and/or magnetic impurities and would be difficult to observe experimentally.

In this paper we show that the pair of MFs at each end of a TR-symmetric Kitaev chain are in fact topologically robust to a large class of perturbations including
mixing by the $S_z=0$ component of the order parameter ($\Delta_{\uparrow\downarrow}=\Delta_{\downarrow\uparrow}$), transverse hopping (for quasi-1D systems), non-magnetic disorder, and also, importantly, to perturbations that explicitly break the TR symmetry such as Zeeman fields/magnetic impurities (in two orthogonal directions in spin-space) and perturbations rendering $|\Delta_{\uparrow\uparrow}| \neq |\Delta_{\downarrow\downarrow}|$ (i.e., mixing by the $S_y=0$ component of the order parameter, $\Delta_{\uparrow\uparrow}=\Delta_{\downarrow\downarrow}=p\Delta_1$). Note that such TR-breaking perturbations are likely to be present in the experiments as the evidence for the possible spin-triplet order in Bechgaard salts is found only in the presence of magnetic fields \cite{Lee_2001,Lee_2003,Shinagawa_2007}. We show that the topological robustness to the TR-breaking perturbations results from a hidden chiral symmetry that places the TR-symmetric Kitaev chain in the BDI topological class with an integer $\mathbb{Z}$ invariant. The integer invariant is equal to the number of MF modes at each end protected by the chiral symmetry. In quasi-1D systems with multiple coupled chains
the $\mathbb{Z}$ invariant can take arbitrary integer values equal to the number of the chains. Our work clarifies the topological properties of the doubled Kitaev chains and the related quasi-1D superconductors with a spin-triplet $p$-wave order parameter.
Additionally, our work shows that the MFs and the resultant zero bias tunneling peak \cite{Sengupta_2001,Law_2009}  and the fractional ac Josephson effect \cite{Kitaev_2001,Kwon_2004} should be topologically robust and experimentally observable in (TMTSF)$_{\rm {2}}$X and Li$_{0.9}$Mo$_6$O$_{17}$.



\begin{figure}
\includegraphics[width=8cm]{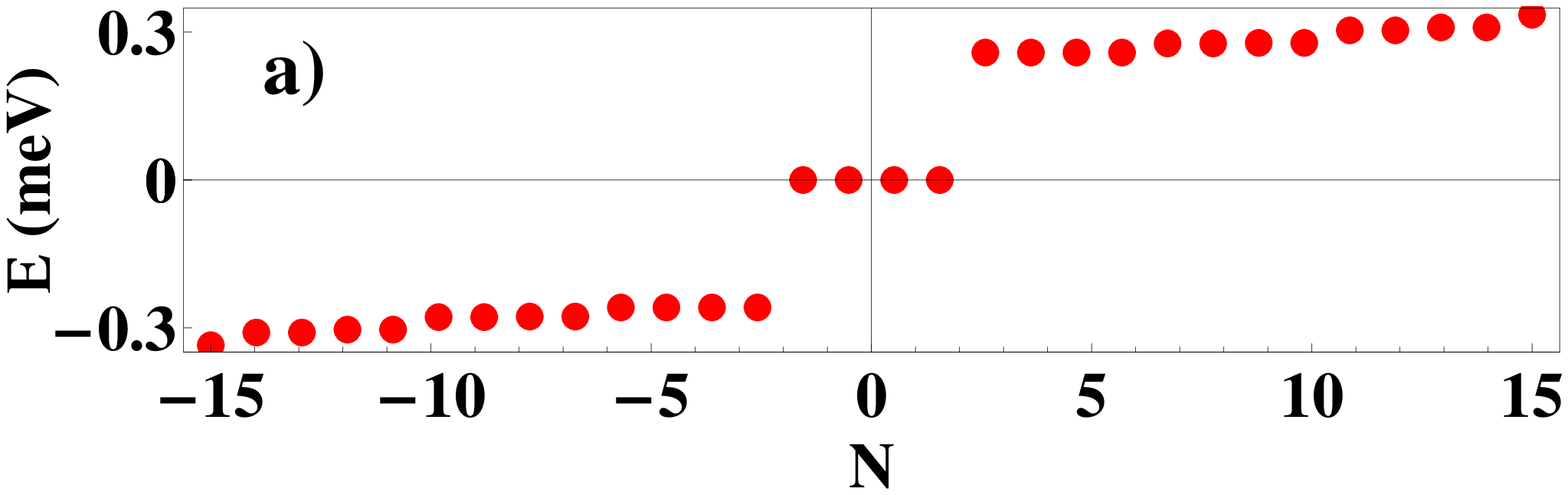}
\includegraphics[width=8cm]{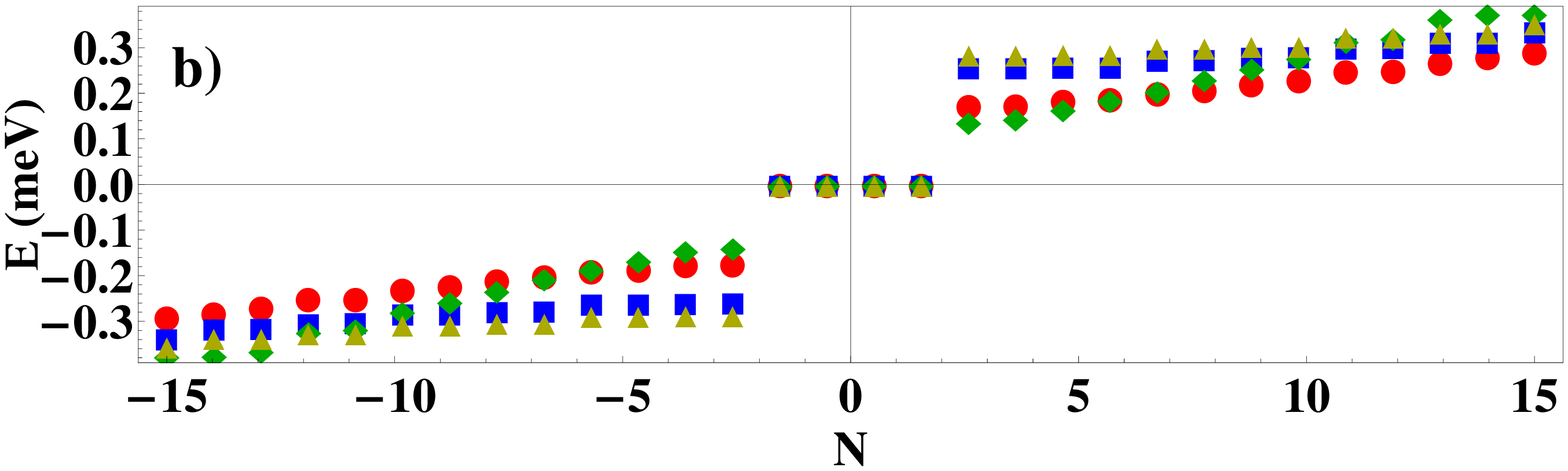}
\includegraphics[width=8cm]{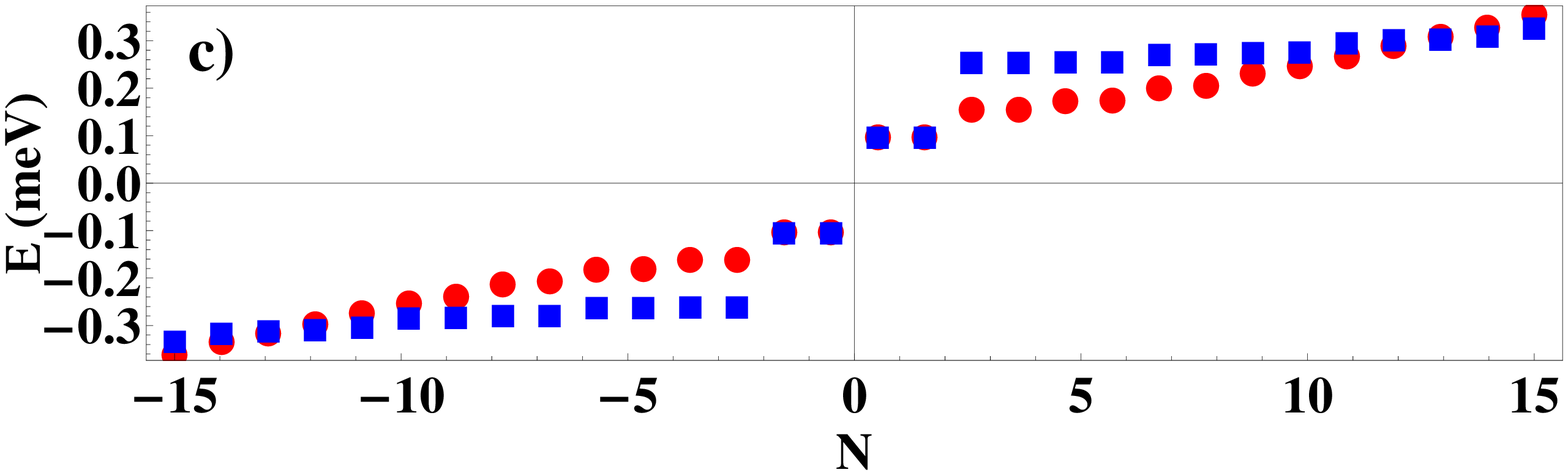}
\caption{\label{fig:LowE1} (Color online) (a) Low energy BdG quasiparticle spectrum for TR-symmetric Kitaev chain corresponding to Eq.~(\ref{H1}) which hosts a pair of Majorana Fermions. The parameters used are as follows: $\Delta_0=0.5$meV, $\mu=0.75$ meV, $N=300$ sites, $a=15$ nm, corresponding to a wire length of $4.5\mu$m and a hopping amplitude of $t=11$meV. (b) The response of the low energy spectrum to the TR-breaking perturbations given in Eq.~(\ref{H2}). We have added bulk Zeeman splitting (red circles, $V_y, V_z=0.4$ meV), magnetic impurity (blue squares) at wire endpoints of magnitude $J_y,J_z=1$ meV, and mixing by the $S_y=0$ ($S_z=0$) component of the order parameter $\Delta_{\uparrow\uparrow}=\Delta_{\downarrow\downarrow}=0.5\Delta_{0}$, green diamonds ($\Delta_{\uparrow\downarrow}=\Delta_{\downarrow\uparrow}=0.5\Delta_{0}$, yellow triangles). (c) Zeeman splitting in the $\hat{x}$ direction ($V_x=0.1$ meV, red circles) and localized at the endpoints ($J_x=0.5$ meV, blue squares) split the MFs to finite energy}
\end{figure}
We begin with the 1D electronic tight binding Hamiltonian for a pair of Kitaev chains in uncoupled spin sectors,
\begin{equation}
H_{1}=\sum_{i,\alpha}(-t c_{i\alpha}^{\dagger}c_{i+1,\alpha}^{  } -\mu c_{i\alpha}^{\dagger}c_{i\alpha}^{ }+\Delta_{\alpha\alpha}c_{i\alpha}^{\dagger}c_{i+1\alpha}^{\dagger}+h.c.),
\label{H1}
\end{equation}
where $i=1,...,N$ represents the lattice sites and $\alpha=\uparrow,\downarrow$ is the spin index. The first term in Eq.~(\ref{H1}) represents the kinetic energy, $\mu$ is the chemical potential and $\Delta_{\alpha\alpha}$ is the ESP $p$-wave superconducting pair potential. For a finite wire we first solve the BdG equations corresponding to Eq.~(\ref{H1}) for the TR-symmetric case, $\Delta_{\uparrow\uparrow}=-\Delta_{\downarrow\downarrow}=p\Delta_0$ (see below for the TR-invariance of the doubled Kitaev chain). As shown in Fig.~(\ref{fig:LowE1}a), the low-energy spectrum for this system contains a total of four zero energy modes (two on each end) separated by a finite gap on each side. The wave functions for the pair of zero modes at each end are such that the corresponding second quantized operators satisfy the Majorana condition, $\gamma_i^{\dagger}=\gamma_i$.

We examine the stability of the MFs against the TR-breaking perturbation Hamiltonian $H_2$,
\begin{eqnarray}
\label{H2}
H_{2}&=&\sum_{i,\alpha,\alpha^{'}}[({\vec{V}\cdot\vec{\sigma}})_{\alpha,\alpha^{'}} c_{i\alpha}^{\dagger}c_{i\alpha^{'}}+J({\vec{S_i}\cdot\vec{\sigma}})_{\alpha,\alpha^{'}} c_{i\alpha}^{\dagger}c_{i\alpha^{'}} \nonumber\\
&+&\Delta_{1}(c_{i\uparrow}^{\dagger}c_{i+1\uparrow}^{\dagger}+c_{i\downarrow}^{\dagger}c_{i+1\downarrow}^{\dagger})+h.c].
\end{eqnarray}
 The first term in Eq.~(\ref{H2}) represents an applied Zeeman field $\vec{V}=(V_{x},V_{y},V_{z})$, the second term represents magnetic impurities localized at site $i$ with
 spin $\vec{S}_i$ and coupling constant $J$, and the third term $\Delta_1$ adds an $S_y=0$ component to the triplet order parameter. All three terms break the TR symmetry (note that the term $\Delta_1$, added to a state with $S_x=0$, makes the magnitudes of the order parameter in the two spin sectors unequal, $|\Delta_{\uparrow\uparrow}| \neq |\Delta_{\downarrow\downarrow}|$). Additionally, we checked the robustness of the MFs to TR-invariant perturbations such as ($\Delta_{\uparrow\downarrow}=\Delta_{\downarrow\uparrow}=p\Delta_2)$, non-magnetic disorder, and, for multiple coupled chains, to hopping in the transverse directions. Since the spin-orbit coupling in the organic superconductors is negligible \cite{Shinagawa_2007} we have not included a Rashba spin-orbit term in the perturbation Hamiltonian.

Fig.~(\ref{fig:LowE1}b) shows the low energy BdG spectrum of the full Hamiltonian $H=H_1+H_2$. The pair of MFs at each end remain protected to perturbations such as Zeeman fields and magnetic impurities along two transverse directions, and also to mixing by the $S_y=0$ and $S_z=0$ components of the triplet order parameter.
In addition we have also found (not shown in Fig.~1) that the MFs are robust to non-magnetic disorder and to transverse hopping (in a quasi-1D system). 
However we find that a Zeeman field along $x$ ($V_x$) splits the MFs to finite energy $\pm \epsilon$ as illustrated in Fig.~(\ref{fig:LowE1}c). Nevertheless, since the MFs are robust to TR-breaking perturbations such as bulk $V_y$ and $V_z$ as well as magnetic impurities and mixing by the $S_y=0$  component of the order parameter, the topological robustness of the MFs cannot be explained by the time reversal symmetry.
To understand the topological properties of the TR-symmetric Kitaev chain we rewrite $H_1$ in Eq.~(\ref{H1}) (for $\Delta_{\uparrow\uparrow}=-\Delta_{\downarrow\downarrow}=p\Delta_0$) as,
as $H_1=\sum_{k}\Psi_{k}^{\dagger} H_{1}(k) \Psi_{k}$ where,
\begin{equation}
H_{1}(k) = (-2t\cos(k)-\mu)\sigma_{0}\tau_{z}+\Delta_0\sin(k)\sigma_{z}\tau_{x},
\label{eq:BDG}
\end{equation}
and the perturbation Hamiltonian $H_2$ as $H_2=\sum_{k}\Psi_{k}^{\dagger} H_{2}(k) \Psi_{k}$ with,
\begin{equation}
H_{2}(k) = V_x\sigma_{x}\tau_{z}+V_{y}\sigma_{y}\tau_{0}+V_z\sigma_{z}\tau_{z}+
 \Delta_1\sin(k)\sigma_0\tau_x,
\label{eq:BDG2}
\end{equation}
where $\sigma_i$ and $\tau_i$ are the Pauli matrices in the spin and the particle-hole spaces, respectively.
Here we have used the coupled spin and particle-hole basis, $\Psi_{k}=(c_{k\uparrow},c_{k\downarrow},c_{-k\uparrow}^{\dagger},c_{-k\downarrow}^{\dagger})^{T}$, and have replaced $k_{x}$ by $k$. Using the time reversal operator $\Theta$ ($\Theta$=$i\sigma_y\tau_0{\cal{K}}$) and the particle-hole operator $\Xi$ ($\Xi=\sigma_0\tau_x{\cal{K}}$) in this basis, where ${\cal{K}}$ is the complex conjugation operator, it is easy to see that $H_1$ is TR-invariant ($\Theta H_1(k) \Theta^{-1}=H_1(-k)$) and has the particle-hole symmetry ($\Xi H_1(k) \Xi^{-1}=-H_1(-k)$). To understand the origin of the chiral symmetry, we introduce the operator ${\cal O}=-i\sigma_z\tau_z{\cal K}$, with ${\cal O}^2=I$, under which the Hamiltonian is invariant (${\cal O} H_1(k){\cal O}^{-1}=H_1(-k)$).
The presence of ${\cal{O}}$ and the particle-hole symmetry $\Xi$ implies that $H_1$ has the chiral symmetry, $\{H,{\cal{S}}\}=0$, where ${\cal{S}}={\cal{O}}\cdot{\Xi}=\sigma_z\tau_y$ is the chirality operator and $\{.\}$ indicates the anti-commutator. The anti-commutation with ${\cal{S}}$ and commutation with ${\cal{O}}$, along with ${\cal{O}}^2=I$, imply that the TR-symmetric
Kitaev model is in the topological class BDI with an integer ($\mathbb{Z}$) winding number invariant $W$ \cite{Schnyder_2008,Ryu_2010,Tewari_PRL_2012,Tewari_PRB_2012}. Note that all the perturbations in Eq.~(\ref{eq:BDG2}), including non-magnetic disorder and transverse hopping, except $V_x$ (and a magnetic impurity polarized along $x$), respect the chiral symmetry ${\cal S}$. As shown in Fig.~1, a perturbation $V_x$ (and $J_x$), therefore, remove the MFs and create a gap.

To calculate $W$ we off-diagonalize the Hamiltonian in Eq.~(\ref{eq:BDG}),
\begin{equation}
H^{'}=UHU^{\dagger}=\left(\begin{array}{cc}
0 & A(k)\\
A^{\dagger}(-k) & 0
\end{array}\right),
\end{equation}
 and write the determinant of $A(k)$
in a complex polar form, $Det(A)=|Det(A)|e^{i\theta(k)}$. The winding number $W$ is then given by \cite{Tewari_PRL_2012}
$W=1/(2\pi)\int_{-\pi}^{\pi}d\theta(k)$.
\begin{figure}
\includegraphics[width=4cm]{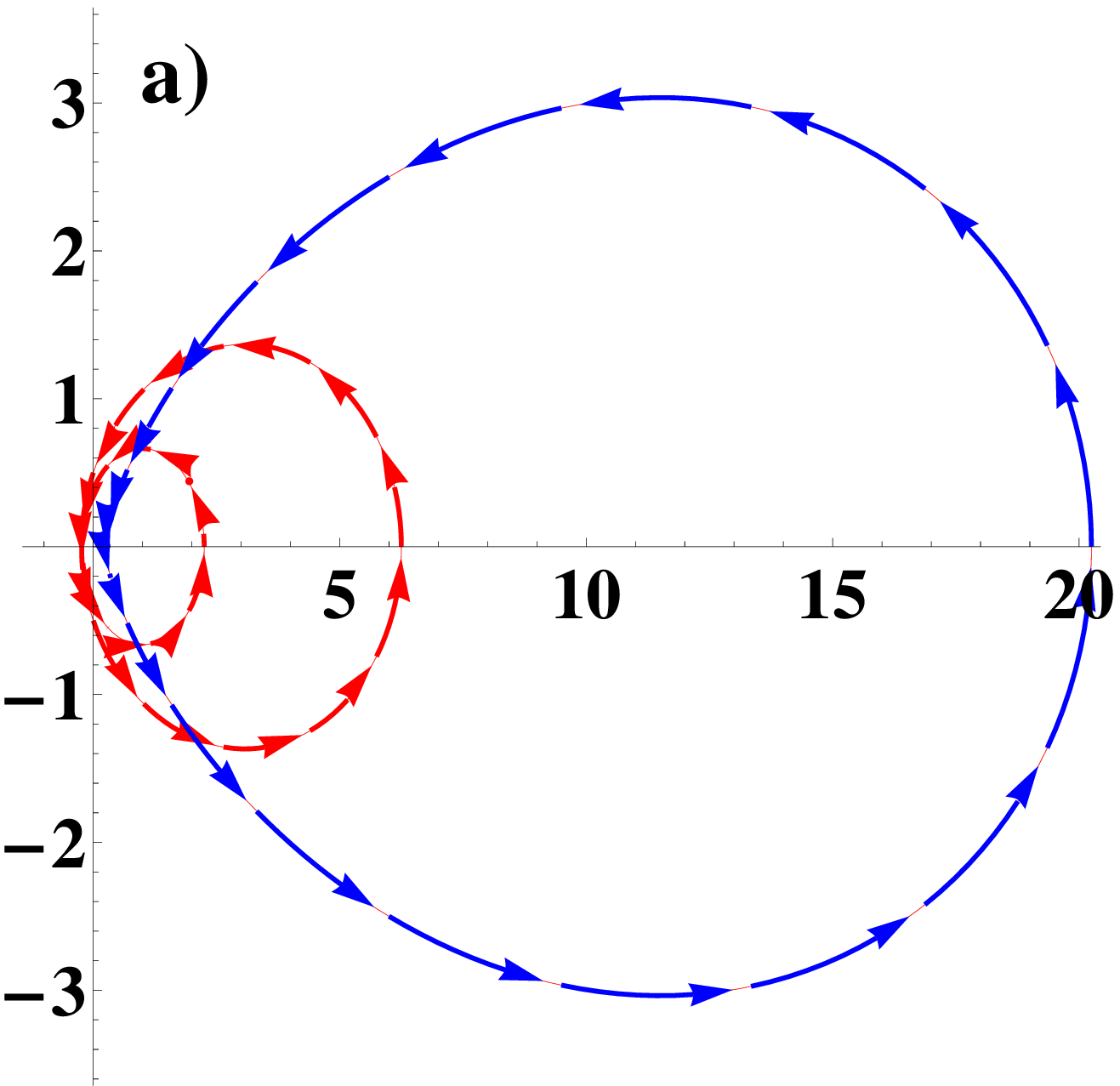}
\includegraphics[width=4cm]{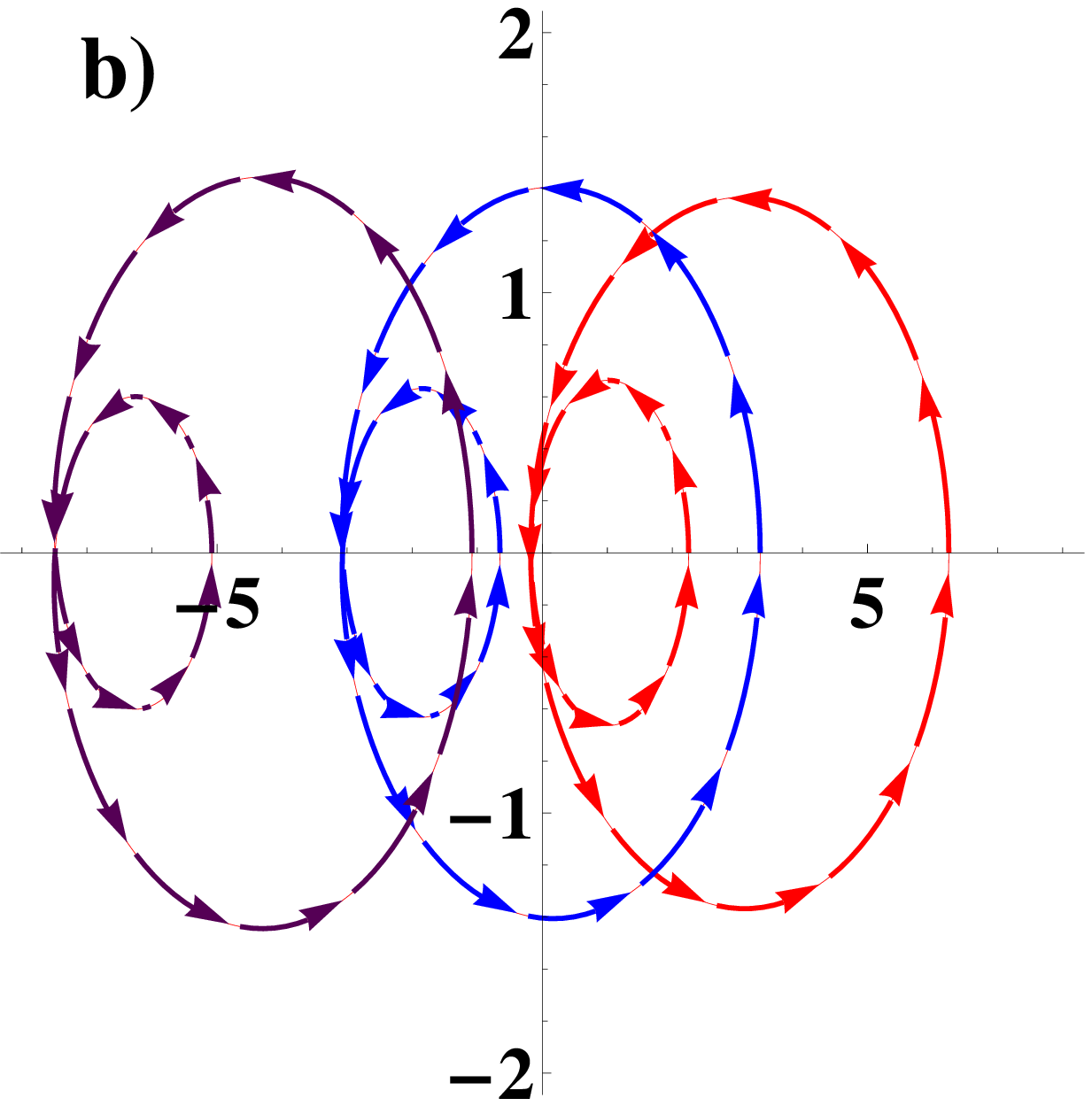}
\caption{\label{fig:Winding} (Color online) Parametric plots of Re$(Det(A(k)))$ and Im$(Det(A(k)))$
as the momentum $k$ is varied through the 1D Brillouin zone from $k=-\pi$ and $\pi$. (a) The winding
of the angle $\theta(k)$ for the TR-symmetric Kitaev chain in the topologically non-trivial (red)
and trivial (blue) phases corresponding to $|\mu|<2t$ and $|\mu|>2t$,
respectively, indicating the existence of 2 and 0 MFs at each end. (b) The winding number in the presence of perturbation ($\Delta_{\uparrow\uparrow}=\Delta_{\downarrow\downarrow}=0.5\Delta_{0}$), and 3 values of the
Zeeman splitting, $V_y,V_z=0.5\mu$ (red), $1.5\mu$ (blue), and $3.3\mu$ (purple). 
With increasing Zeeman fields, the Fermi surfaces disappear in turn and $W$ decreases from 2 (red) to 1 (blue) to 0 (purple) indicating the corresponding disappearance of the MFs.}
\end{figure}
In Fig.~(\ref{fig:Winding}a), we show the winding number in the topologically non-trivial and trivial phases of the TR-symmetric Kitaev chain. The
winding number is 2 (0) in the topologically non-trivial (trivial) phases indicating the existence of 2 (0) topologically protected MFs on a given end.
In Fig.~(\ref{fig:Winding}b) we show that the winding number is 2 even in the presence of the perturbations $V_y,V_z$ and $\Delta_{\uparrow\uparrow} = \Delta_{\downarrow\downarrow}\neq 0$ (red curve). This explains the topological robustness of the pair of MFs on a given end of the finite wire (Fig.~(1)). In Fig.~(2b), blue and purple curves indicate the evolution of $W$ with increasing Zeeman fields $V_y,V_z$. Even though the chiral symmetry is still unbroken, the Zeeman field can change the value of $W$ from $2 \rightarrow 1 \rightarrow 0$, indicating a corresponding decrease of the number of MFs on a given end. Physically, the Zeeman field reduces the number of MFs by 1 by removing the Fermi surfaces in turn. 

\begin{figure} [t!]
\includegraphics[width=4cm]{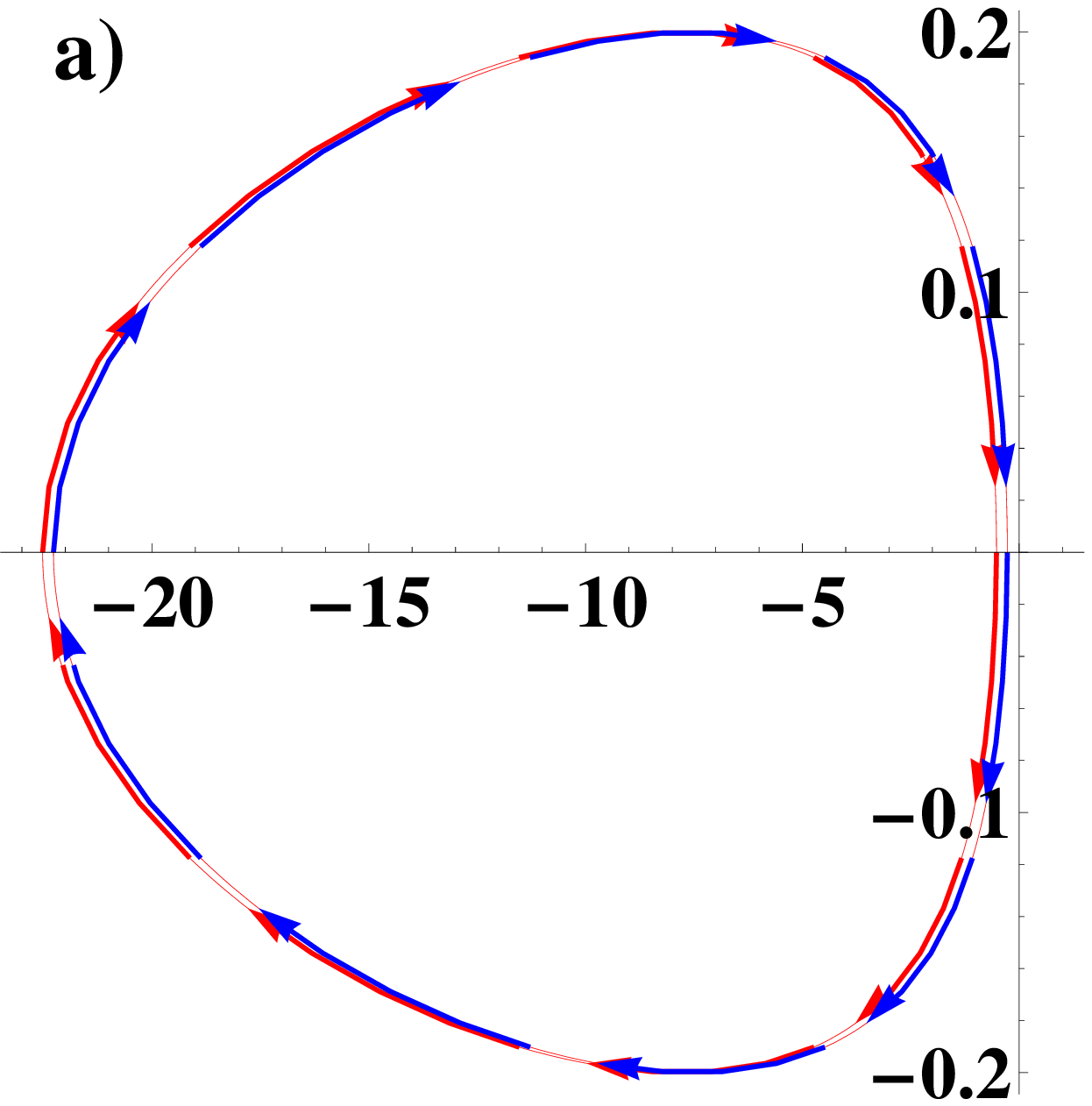}
\includegraphics[width=4cm]{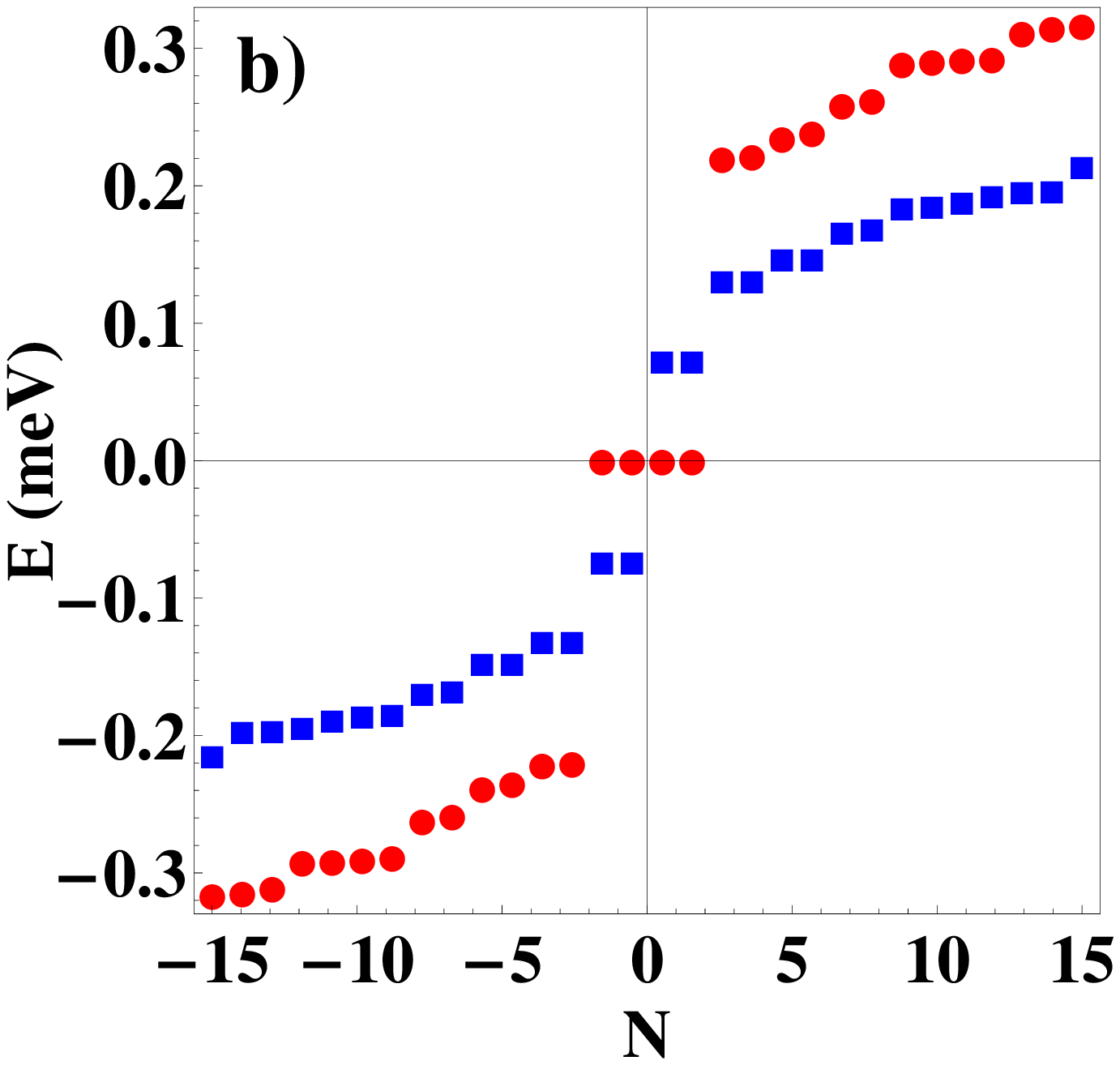}
\includegraphics[width=4cm]{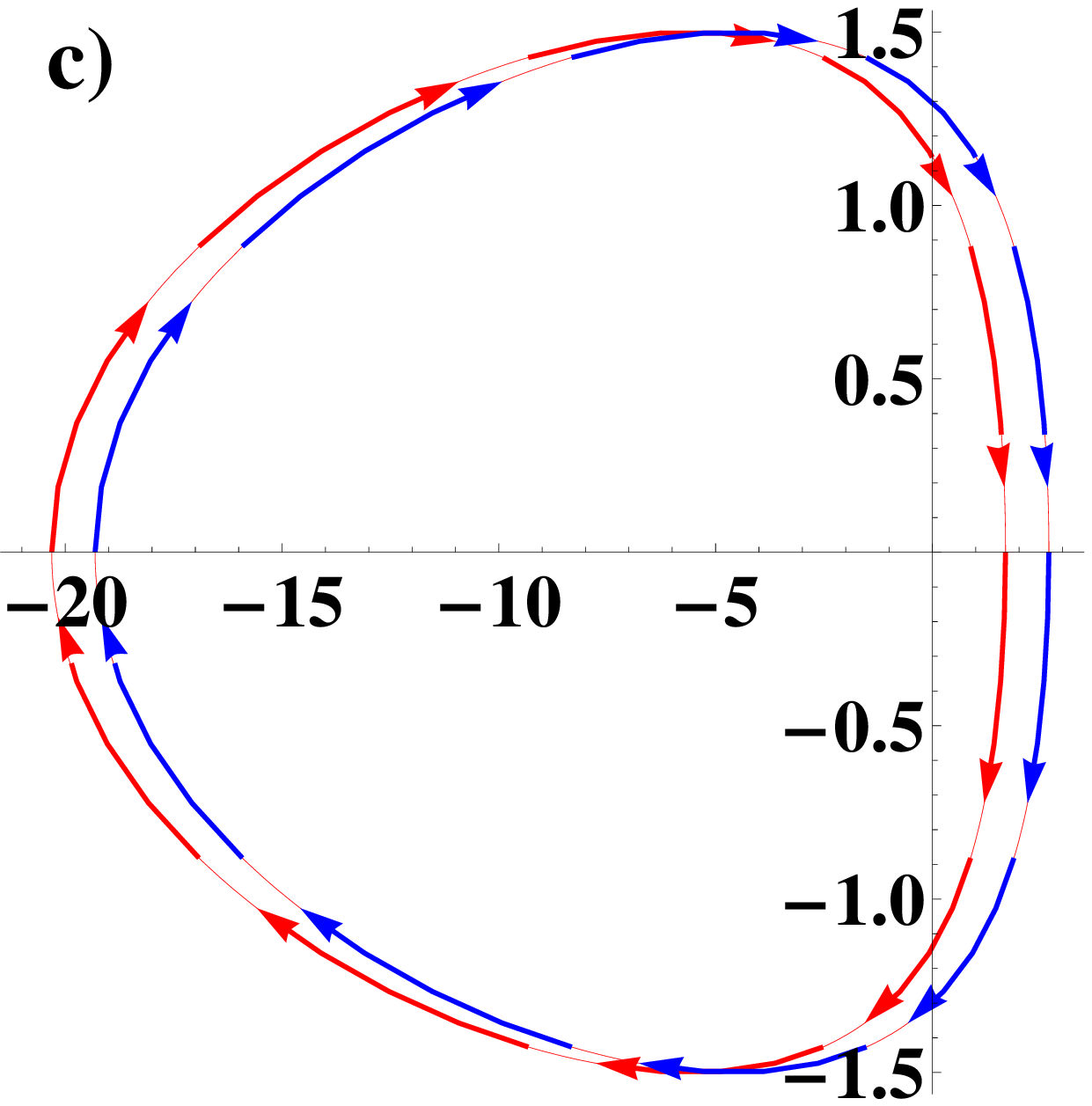}
\includegraphics[width=4cm]{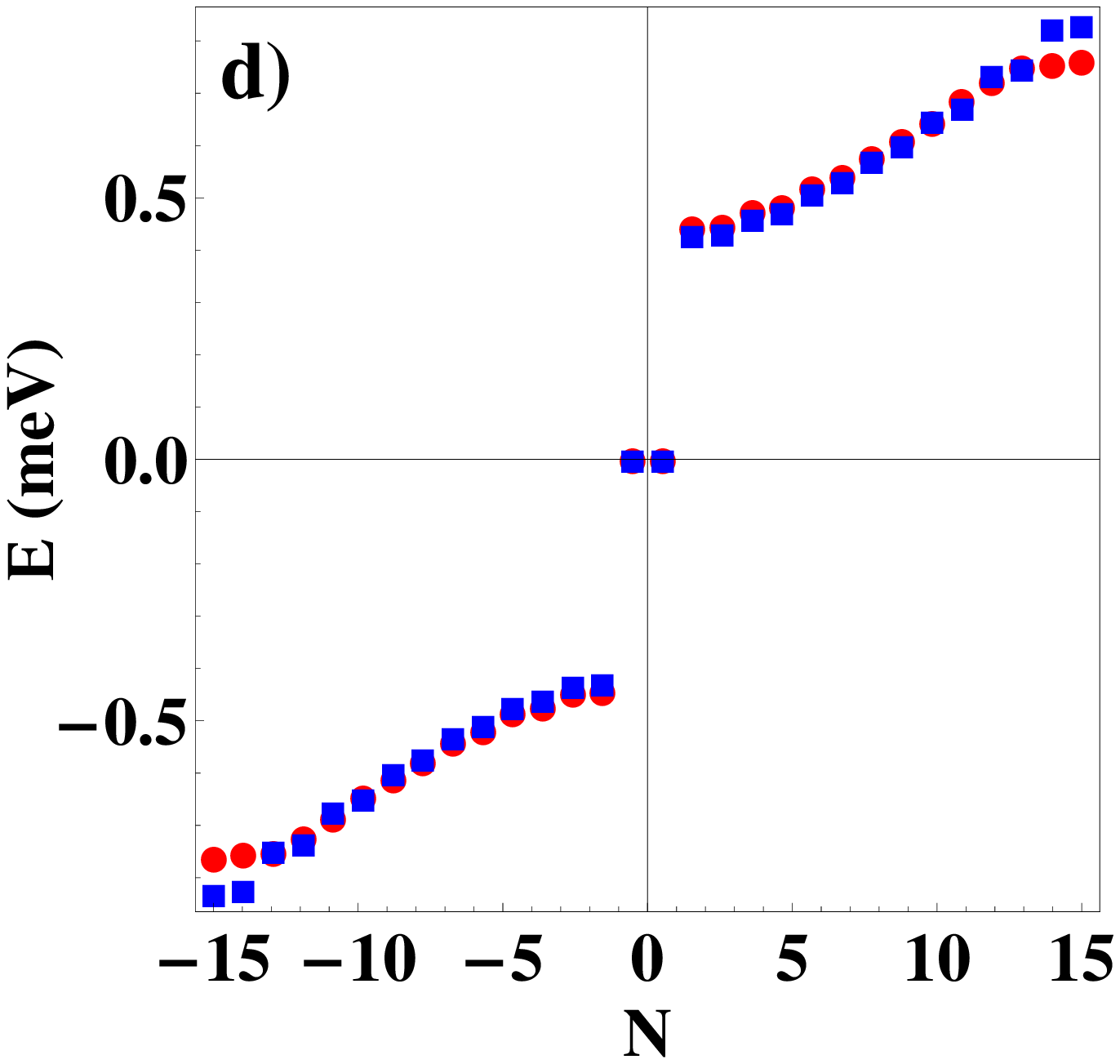}
\caption{\label{fig:Winding2}  (a) The winding of the angle $\theta(k)$ for the ($S_y=0$) Kitaev chain with $\Delta_{\uparrow\uparrow}=\Delta_{\downarrow\downarrow}$ with the two spin sectors uncoupled (red) and coupled by a Zeeman splitting $V_y=0.25$ meV (blue). In both cases $W=0$. (b) Low energy BdG spectrum for the uncoupled spins (red circles) shows 2 \textit{topologically unprotected} MFs at each end.
 Since the MFs are unprotected (see text), even a chiral symmetric perturbation $V_y=0.25$ meV splits them to finite energy (blue squares). (c) Increasing the Zeeman splitting in the $\hat{z}$ direction ($V_{z}>\mu$) drives the system through a topological phase transition into a phase with winding number $W=1$. We show the winding for uncoupled spins (red) and with additional Zeeman splitting $V_y=0.25$ meV (blue). (d) Low energy BdG spectrum in the $W=1$ regime for uncoupled spins (red circles) and in the presence of $V_y$ (blue squares) illustrating the topological protection of the MF.}
\end{figure}

Next we consider the Hamiltonian in Eq.~(\ref{H1}) with $\Delta_{\uparrow\uparrow}=\Delta_{\downarrow\downarrow}$ (ESP phase with $S_y=0$).
In this phase, as we show in Fig.~(3a), the winding number $W=0$ (red curve). Although $W=0$, since the two spin sectors are uncoupled and each sector hosts a single MF, the system has
a total of 2 MFs on a single end. However, in this case, the pair of MFs are \textit{accidental} with respect to the chiral symmetry ${\cal{S}}$, meaning that they are \textit{not} topologically protected by ${\cal S}$. We see the absence of the topological protection in Fig.~(3b) where a small Zeeman field $V_y$, even though it respects the symmetry under ${\cal S}$ (see Eq.~(\ref{eq:BDG2})), still removes the MFs from the chain ends.
Fig.~(3c,3d) illustrate the fact that in the $S_y=0$ phase,
a field $V_z > \mu$ can remove one of the Fermi surfaces, resulting in the winding number increasing from 0 to 1 (Fig.~(3c)), which is then protected by the chiral symmetry ${\cal S}$. We note in passing that the pair of MFs in the $S_y=0$ phase are also protected to $V_x, V_z$, but this protection
is provided by a different chiral symmetry $\tilde{\cal S}=\sigma_0\tau_y$. The topological properties of a general 1D spinful $p$-wave superconductor are beyond the scope of the present paper and will be taken up in a future publication.



\begin{figure}
\label{fig:Transverse}
\includegraphics[width=8cm]{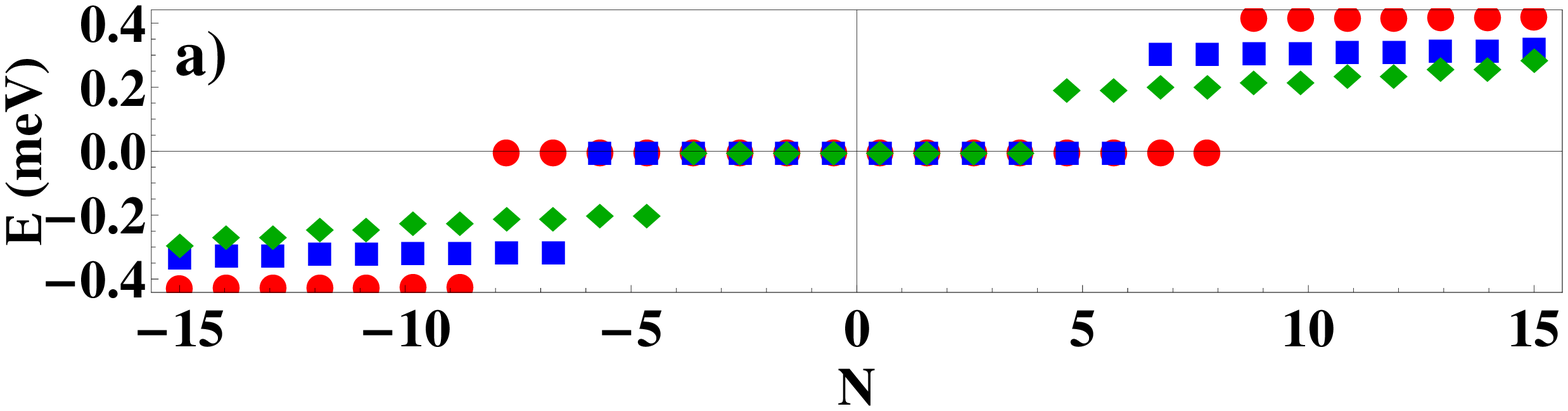}
\includegraphics[width=8cm]{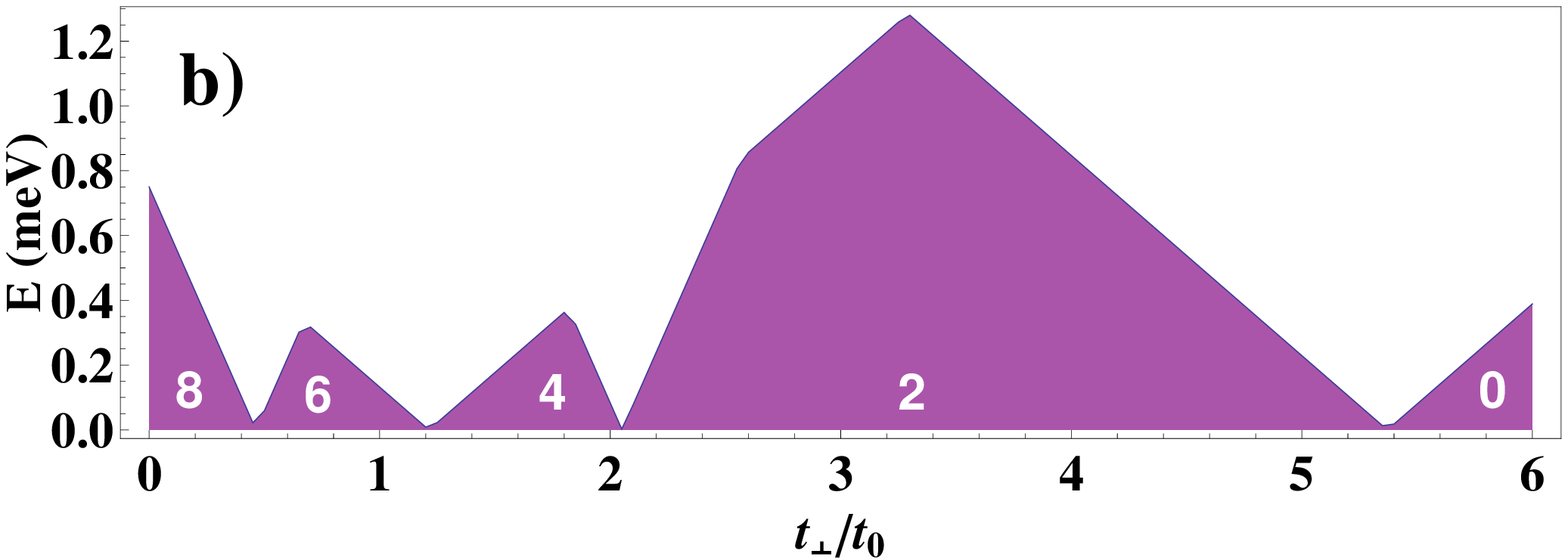}
\includegraphics[width=4cm]{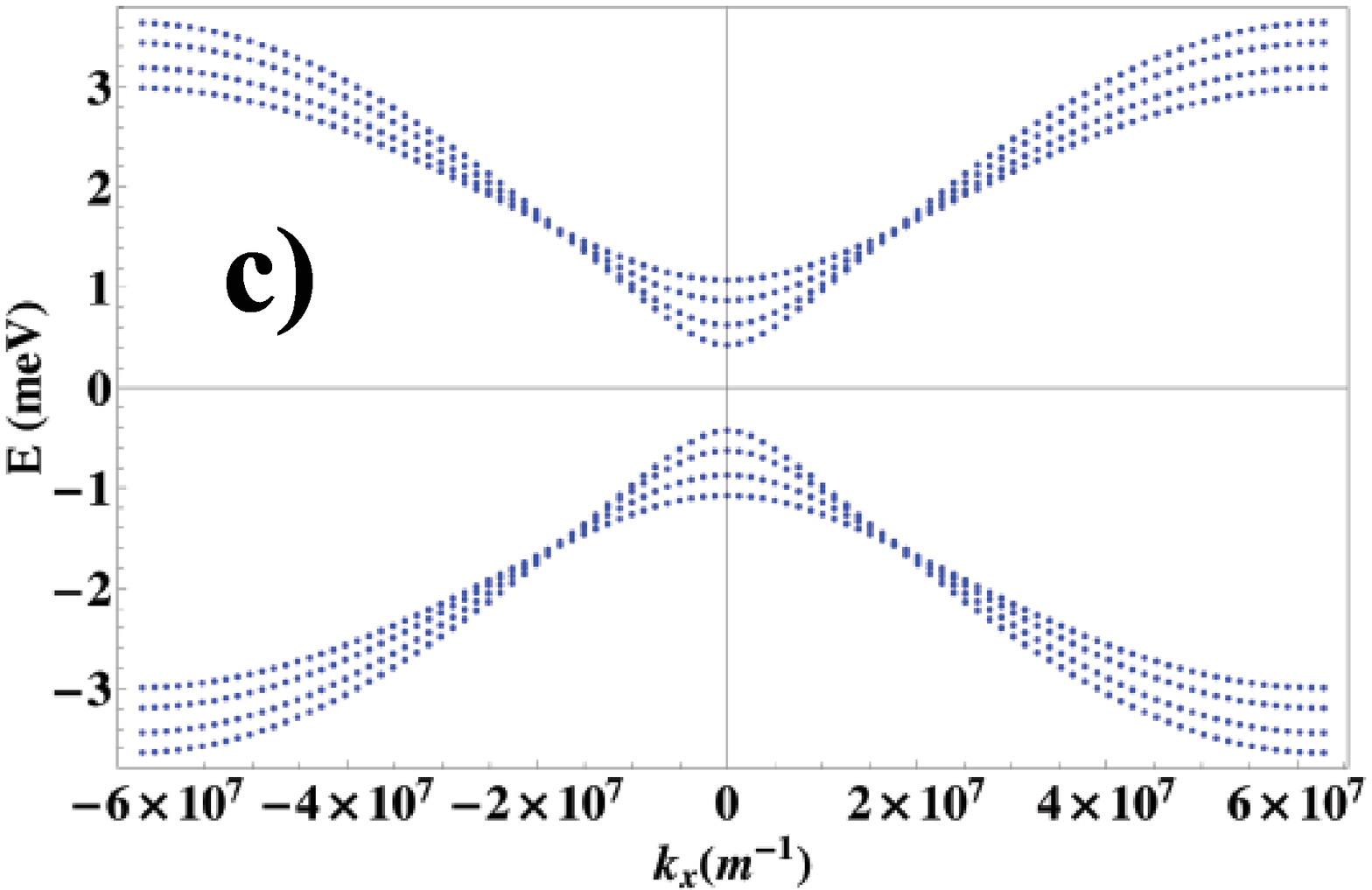} \includegraphics[width=4cm]{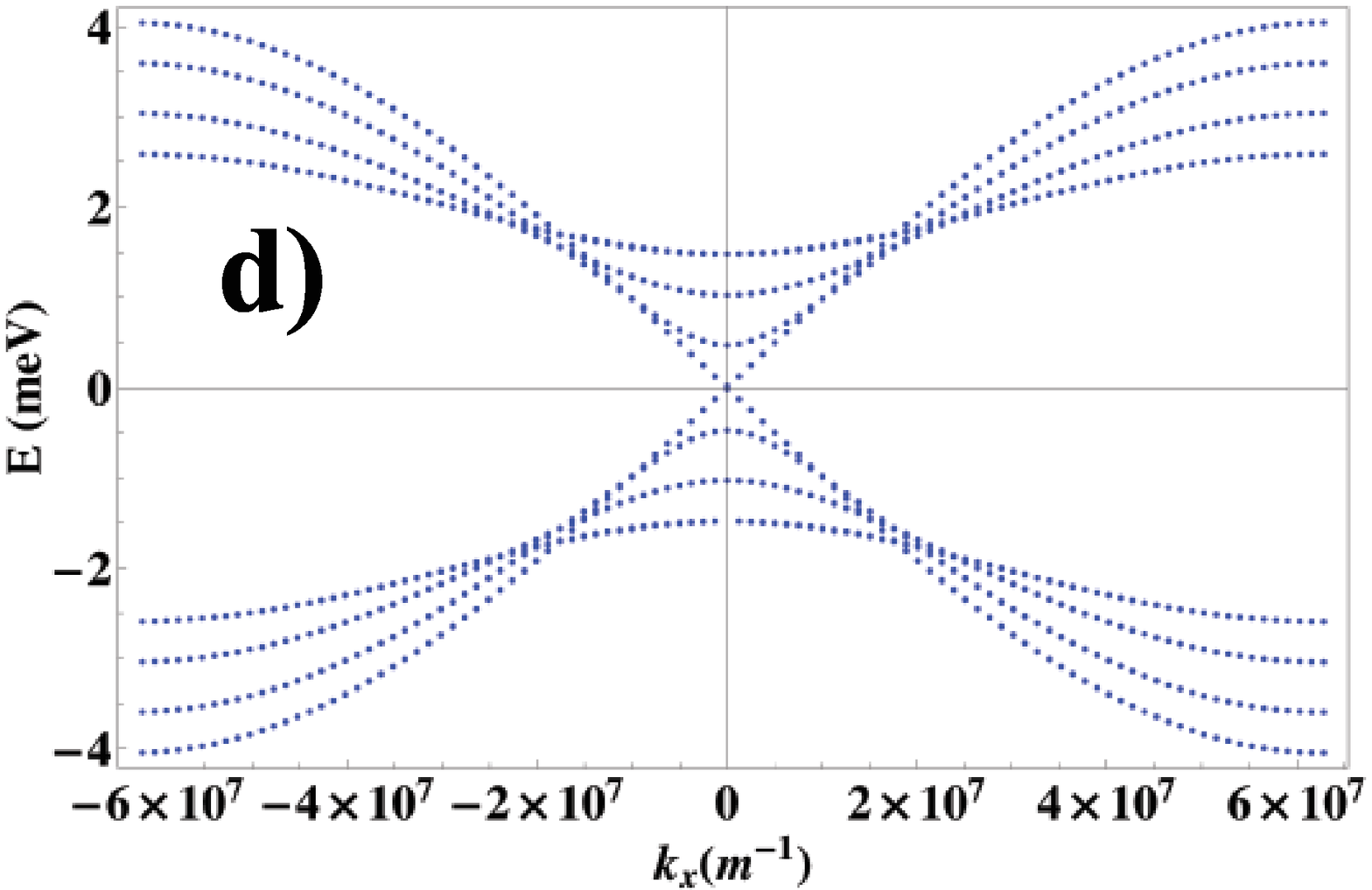}
\caption{(a) Low energy BdG spectrum of four parallel chains coupled by transverse hopping $t_y$. $t_y/t_x=0.2,0.75,1.5$ correspond to 8 (red circles), 6 (blue circles), and 4 (green diamonds) MFs at each end. (b) Quasiparticle gap closing and TQPT (separating phases with different number of MFs) tuned by the transverse hopping.  Panels (c, d) show the bulk energy-momentum dispersion in the gapped regime with 8 MFs and as the bulk gap closes leading to the regime with 6 MFs at each end.}
\end{figure}
To illustrate the
possibility for higher integer values of $W$ (and, correspondingly, higher number of protected MFs per end) we now consider multiple chains
coupled in the transverse directions by hopping parameters $t_y, t_z$. Since in the Bechgaard salts $t_x:t_y:t_z=1:0.1:0.03$ we
consider only the effects of $t_y$.
In the limiting case $t_y \rightarrow 0$
there exist a set of degenerate 1D chains each
hosting 2 MFs at each end.
A small hopping between the chains $t_{y} \ll t_x$
breaks this degeneracy.
Despite the degeneracy breaking,
the additional terms in the Hamiltonian do not break the chiral symmetry and the multi-chain problem is still described by the winding number $W$ (suitably defined for a larger dimensional $H$).
If $t_y \ll t_x$ between
$N$ conducting chains, such that the confinement energy is much smaller
than the chemical potential, the system will accommodate $2N$ sets
of Majorana fermions. As the strength of $t_{y}$
increases the confinement energy lifts the energy of the
 bands above $\mu$ such that they are no
longer occupied. Thus the number of MFs localized at an end (and the value of $W$) goes down in pairs.
As shown in Fig.~(4), at each jump of the value of $W$, the superconducting quasiparticle gap closes and the system passes through
a topological quantum phase transition (TQPT).

The MFs in 1D chiral symmetric topological superconductors can be probed by differential tunneling conductance from the ends \cite{Diez_2012}.
\begin{figure}
\includegraphics[width=4cm]{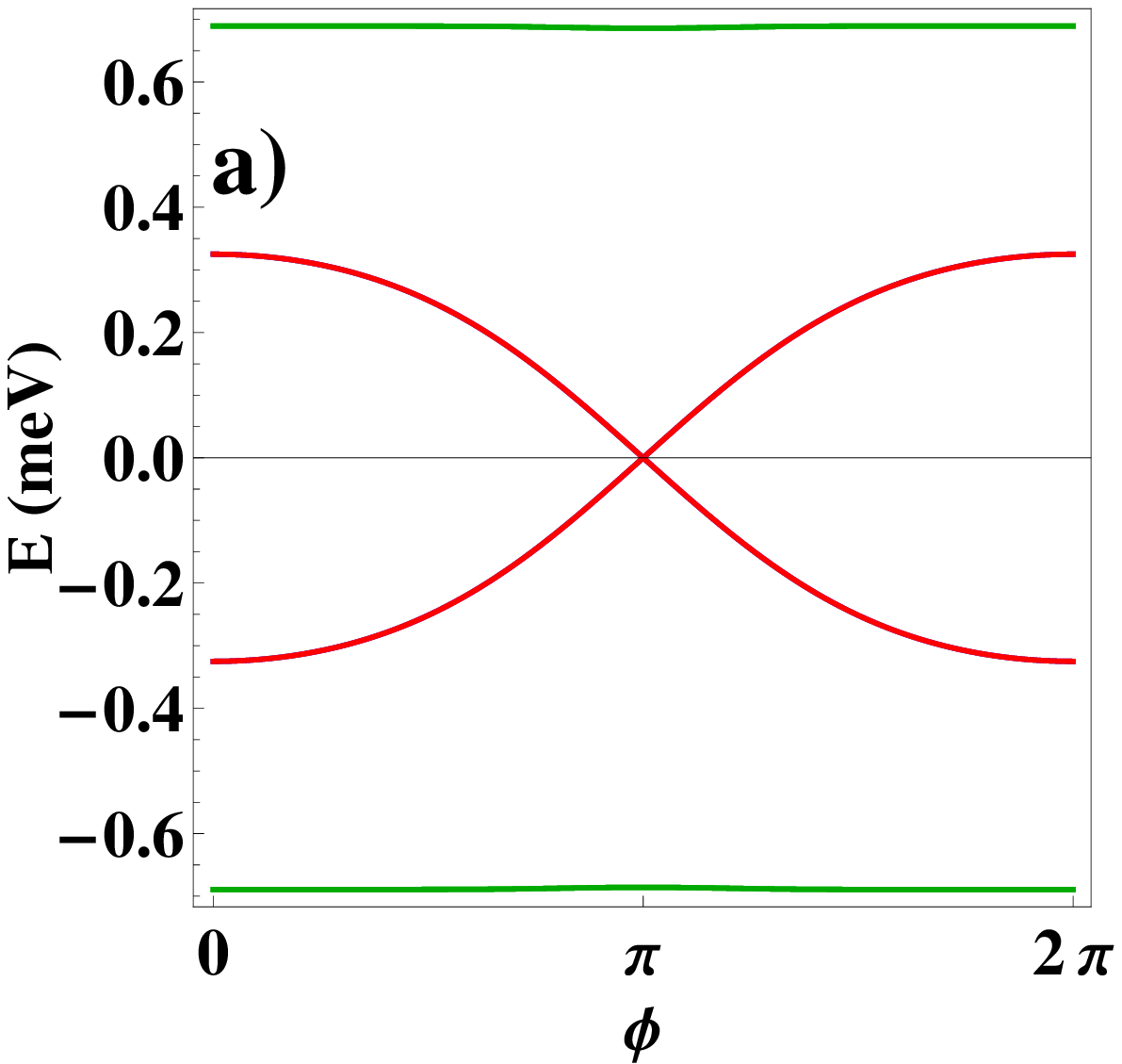} \includegraphics[width=4cm]{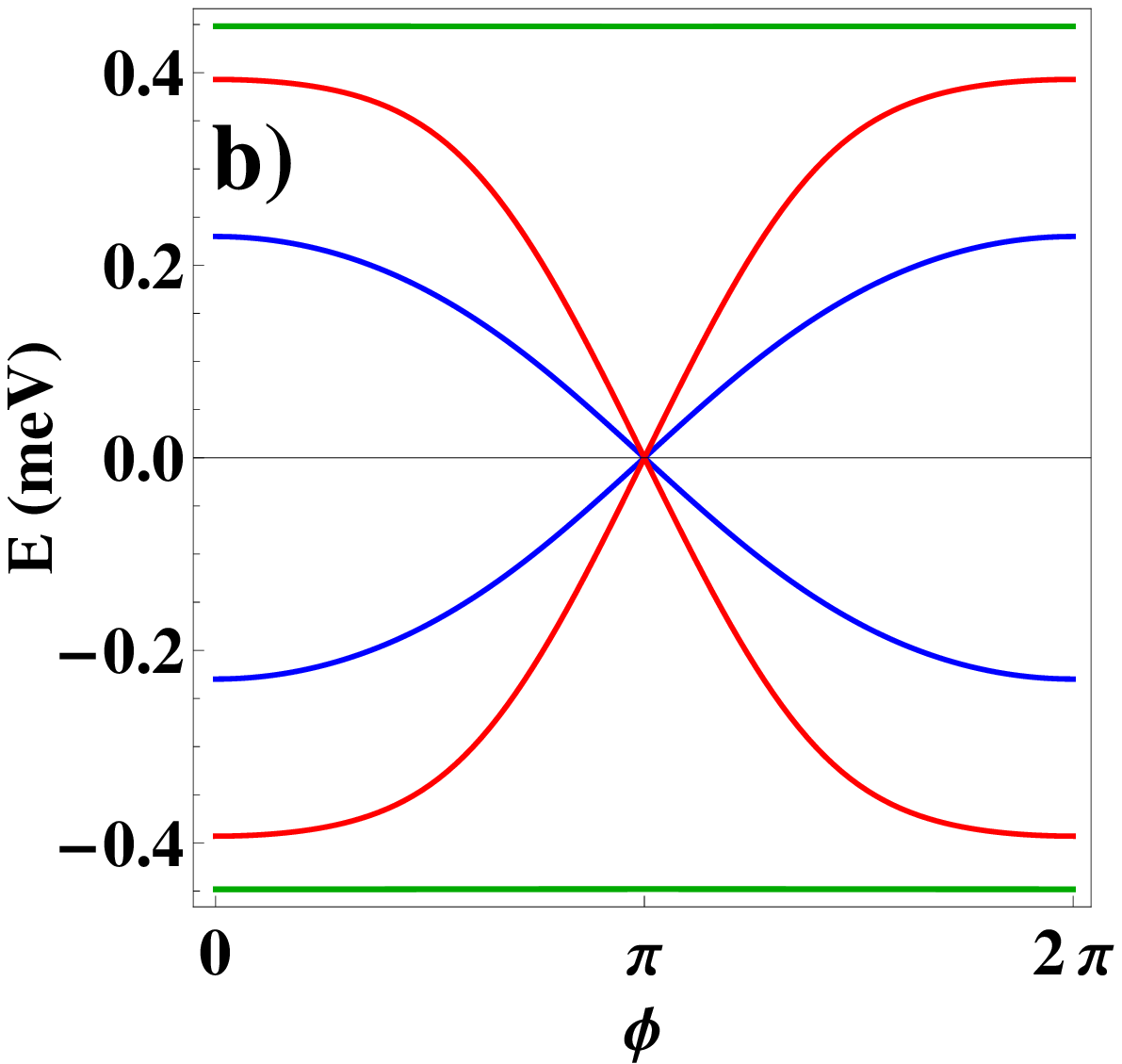}
\includegraphics[width=4cm]{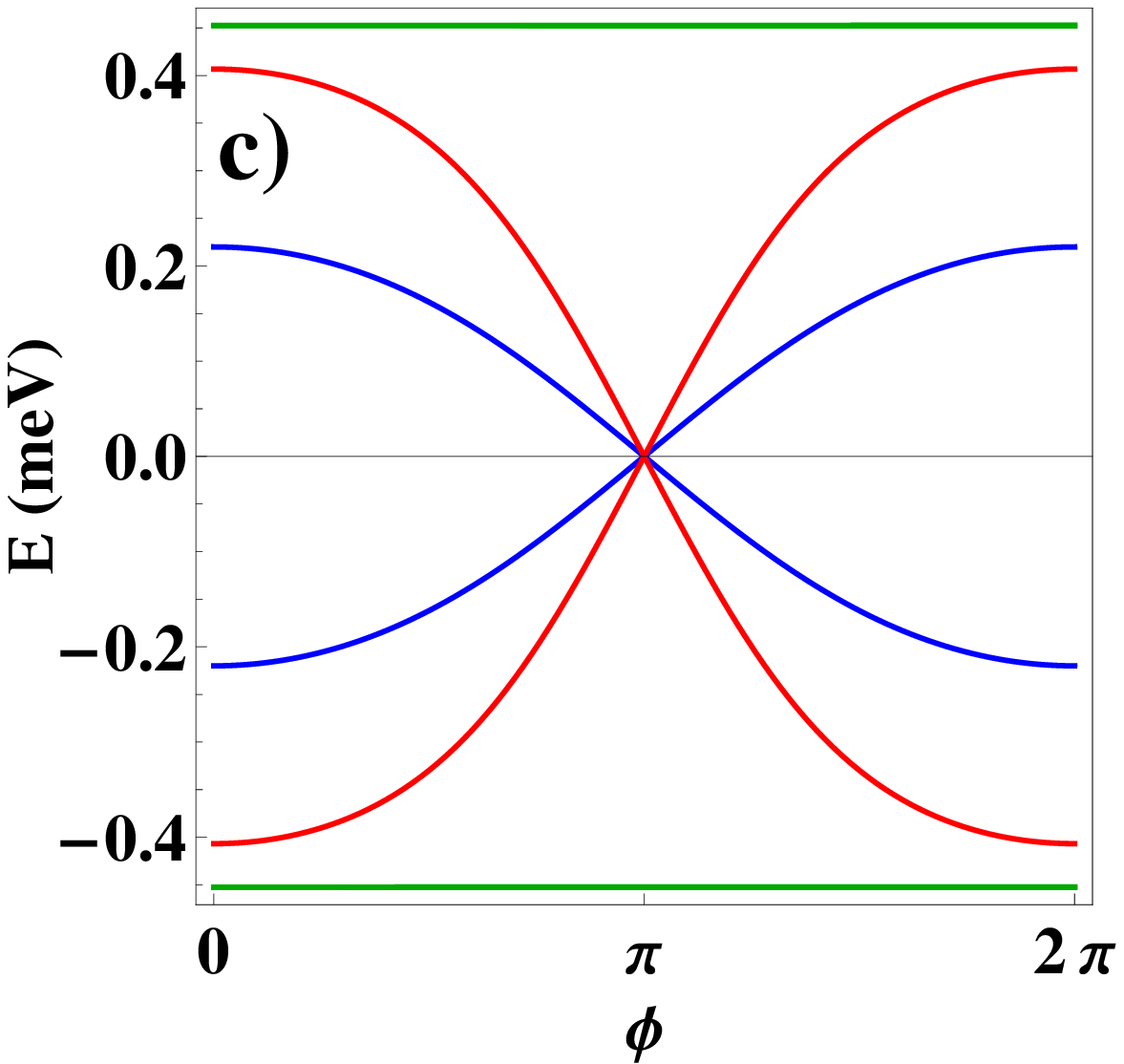} \includegraphics[width=4cm]{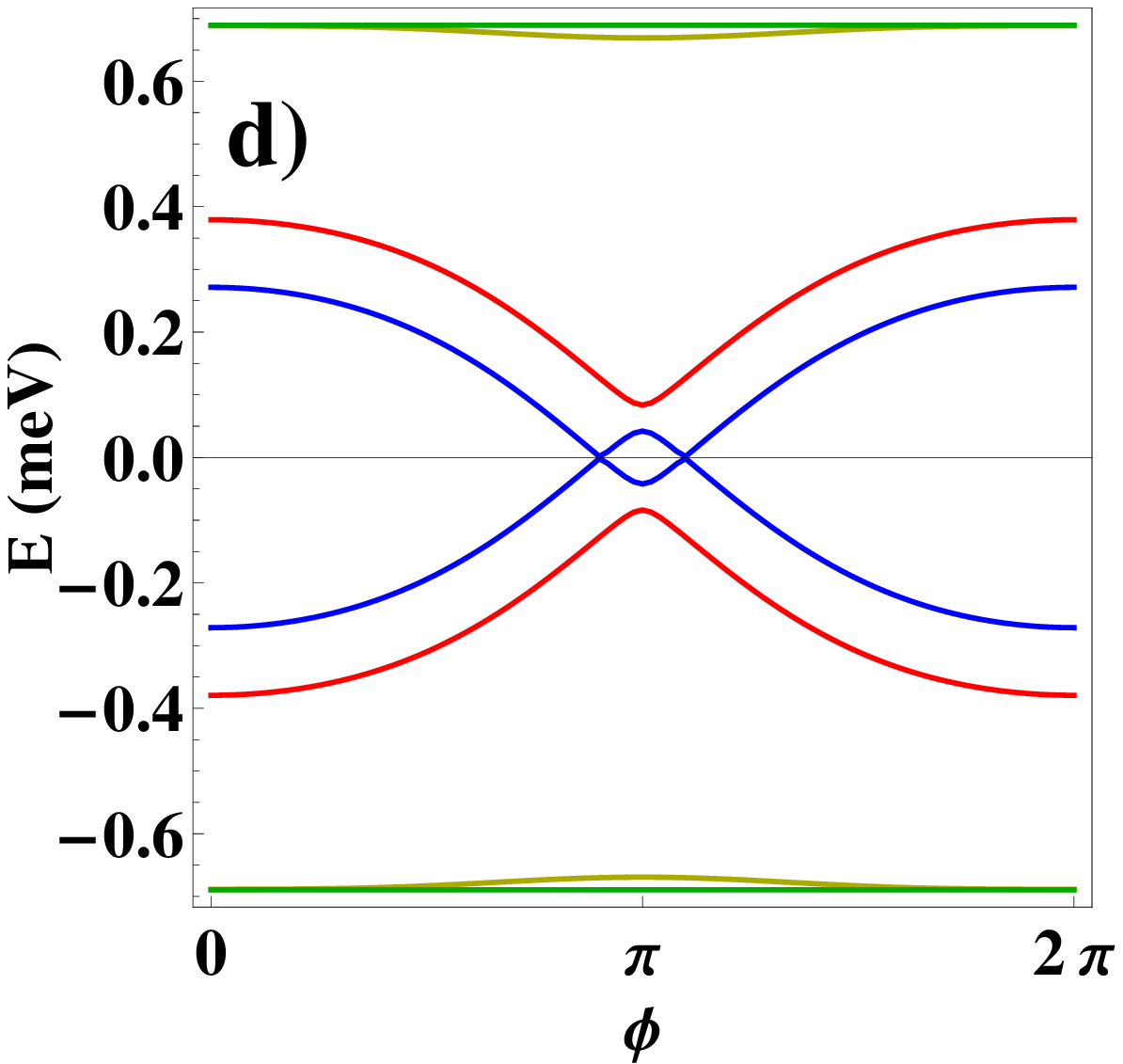}
\caption{\label{fig:ABS} (a) Low energy ABS spectrum as a function of $\phi$ for TR-symmetric Kitaev chain with parameters as in Fig. 1. Each red curve is twofold degenerate. (b) TR-breaking bulk Zeeman fields $V_{y},V_{z}=.25$ meV lift the degeneracy but preserves the $4\pi$ periodicity of the spectrum. (c) Adding the order parameter component with $\Delta_{\uparrow\uparrow}=\Delta_{\downarrow\downarrow}=0.5\Delta_0$, although it breaks the TR symmetry, preserves the $4\pi$ periodicity (d) The spectrum with $V_x=1$ meV added to the junction breaks chiral symmetry and results in a conventional $2\pi$ Josephson effect.}
\end{figure}
In a butt-to-butt Josephson set up between two 1D chiral TS there exists a stable Majorana quartet (two MFs on each side of the junction). The energy levels of the junction plotted as a function of the phase difference $\phi$ is $4\pi$ periodic \cite{Kwon_2004}, giving rise to a
$4\pi$ periodic Josephson effect (see Fig.~(5a)). As shown in Figs.~(5b, 5c), the topological robustness due to chiral symmetry ensures that the single crossing of the
$E$ vs. $\phi$ curves at $\phi=\pi$ is stable to perturbations including those breaking TR symmetry.
 In Fig.~(5d) we show that the $4\pi$ periodicity of the curves is
broken only by adding a Zeeman field $V_x$ that breaks the chiral symmetry ${\cal{S}}$.

In summary we show that the pair of MFs at each end of a 1D
TR symmetric Kitaev $p$-wave chain are topologically robust
to a large number of perturbations including those breaking TR symmetry. We identify the appropriate topological class to be
BDI with an integer ($\mathbb{Z}$) invariant the value of which gives the  number of topologically protected MFs at each end.
In addition to the topological properties of the TR-symmetric Kitaev chains, our results establish the organic superconductors (TMTSF)$_{\rm {2}}$X (X=PF$_6$, ClO$_4$) and Li$_{0.9}$Mo$_6$O$_{17}$, which have been proposed
  \cite{Lee_2001,Lee_2003,Shinagawa_2007} to be quasi-1D equal-spin-pairing $p$-wave superconductors,
  as
  suitable platforms for experimental studies of MFs.

  We thank S. E. Brown, J. D. Sau, K. Sengupta, and E. Ardonne for useful discussions. This work is supported by NSF Grant No. PHY-1104527 and AFOSR (FA9550-13-1-0045).




\begin{thebibliography}{10}

\bibitem{Read_Green_2000} N. Read and D. Green, Phys. Rev. B \textbf{61}, 10267 (2000)

\bibitem{Kitaev_2001} A.Y. Kitaev, Physics-Uspekhi \textbf{44}, 131 (2001).

\bibitem{Nayak_2008} C. Nayak, S. H. Simon, A. Stern, M. Freedman, S. Das Sarma,
	Rev. Mod. Phys. {\bf 80}, 1083 (2008)


\bibitem{Fu_2008} L. Fu, C. L. Kane, Phys. Rev. Lett. \textbf{100}, 096407 (2008)

\bibitem{Zhang_2008} C. W. Zhang, S. Tewari, R. M. Lutchyn, S. Das Sarma, Phys. Rev. Lett. \textbf{101}, 160401 (2008).

\bibitem{Sato_2009} M. Sato, Y. Takahashi, S. Fujimoto, Phys. Rev. Lett. \textbf{103}, 020401 (2009).


\bibitem{Sau} J. D. Sau, R. M. Lutchyn, S. Tewari, and S. Das Sarma, Phys. Rev. Lett. 104, 040502 (2010)

\bibitem{Long-PRB} J. D. Sau, S. Tewari, R. M. Lutchyn, T. D. Stanescu, and S. Das Sarma, Phys. Rev. B 82, 214509 (2010)

\bibitem{Roman} R. M. Lutchyn, J. D. Sau, and S. Das Sarma, Phys. Rev. Lett. 105, 077001 (2010)

\bibitem{Oreg} Y. Oreg, G. Refael, and F. von Oppen, Phys. Rev. Lett.
105, 177002 (2010).

\bibitem{Mourik_2012}V. Mourik, K. Zuo, S. M. Frolov, S.
R. Plissard, E. P. A. M. Bakkers, L. P. Kouwenhoven, Science 336 1003
(2012)

\bibitem{Rokhinson_2012}Leonid P. Rokhinson, Xinyu Liu \& Jacek K. Furdyna,
Nature Physics 8, 795\textendash{}799 (2012)



\bibitem{Deng_2012}M. T. Deng, C. L. Yu, G. Y. Huang, M. Larsson, P.
Caroff and H. Q. Xu, Nano Lett. 12 6414, (2012)

\bibitem{Das_2012}Anindya Das, Yuval Ronen, Yonatan Most, Yuval Oreg,
Moty Heiblum \& Hadas Shtrikman, Nature Phys. 8, 887 (2012)

\bibitem{Churchill_2013} H. O. H. Churchill, V. Fatemi, K. Grove-Rasmussen, M. T. Deng, P. Caroff,
H. Q. Xu, C. M. Marcus, Phys. Rev. B \textbf{87}, 241401(R) (2013)

\bibitem{Finck_2013} A. D. K. Finck, D. J. Van Harlingen, P. K. Mohseni, K. Jung, X. Li, Phys. Rev. Lett. \textbf{110}, 126406 (2013)

\bibitem{Alicea_2012} J. Alicea, Rep. Prog. Phys. \textbf{75}, 076501 (2012)

\bibitem{Leijnse_2012} M. Leijnse, K. Flensberg, Semicond. Sci. Technol. \textbf{27}, 124003 (2012)

\bibitem{Beenakker_2013} C. W. J. Beenakker, Annu. Rev. Con. Mat. Phys. \textbf{4}, 113 (2013)

\bibitem{Stanescu_2013} T. D. Stanescu, S. Tewari, J. Phys. Condens. Matter \textbf{25}, 233201 (2013)



\bibitem{Schnyder_2008}A. P. Schnyder, S. Ryu, A. Furusaki, and A. W. W. Ludwig, Phys. Rev. B \textbf{78} 195125 (2008);
A. P. Schnyder, S. Ryu, A. Furusaki, and A. W. W. Ludwig,  AIP Conf. Proc. \textbf{1134} 10 (2009).

\bibitem{Kitaev_2009} A. Yu Kitaev AIP Conf. Proc. \textbf{1134} 22 (2009).




\bibitem{Ryu_2010} S. Ryu, A. Schnyder, A. Furusaki, A. W. W. Ludwig, New J. Phys. \textbf{12}, 065010 (2010)

 \bibitem{Lee_2001} I. J. Lee, S. E. Brown, W. G. Clark, M. J. Strouse, M. J.Naughton, W. Kang, and P. M. Chaikin, Phys. Rev. Lett.
\textbf{88}, 017004 (2001)

\bibitem{Lee_2003} I. J. Lee, D. S. Chow, W. G. Clark, M. J. Strouse, M. J.
Naughton, P. M. Chaikin, and S. E. Brown, Phys. Rev. B
\textbf{68}, 092510 (2003)

\bibitem{Shinagawa_2007} J. Shinagawa, Y. Kurosaki, F. Zhang, C. Parker, S. E. Brown, D. Jérome, J. B. Christensen, and K. Bechgaard
Phys. Rev. Lett. \textbf{98}, 147002 (2007)

\bibitem{Lebed_2000} A. G. Lebed, K. Machida, M. Ozaki, Phys. Rev. B \textbf{62}, R795 (2000)

\bibitem{Ardavan_2012} A. Ardavan, S. Brown, S. Kagoshima, K. Kanoda,
K. Kuroki, H. Mori, M. Ogata, S. Uji, and J. Wosnitza, J. Phys. Soc.
Jpn. \textbf{81}, 011004 (2012).

\bibitem{Jerome_2008} C. Bourbonnais and D. Jerome, \textit{Physics of Organic Superconductors and Conductors}(Springer, Berlin, 2008).

\bibitem{Thomale_2013} W. Cho, R. Thomale, S. Raghu, and S. A. Kivelson, Phys. Rev. B \textbf{88}, 064505 (2013).


\bibitem{Lebed_2013} A. G. Lebed and O. Sepper
Phys. Rev. B \textbf{87}, 100511 (2013)

\bibitem{Sengupta_2001} K. Sengupta, I. Zutic, H.-J. Kwon, V.
M. Yakovenko, and S. Das Sarma, Phys. Rev. B \textbf{63}, 144531 (2001)


\bibitem{Law_2009} K. T. Law, P. A. Lee, and T. K. Ng,
	Phys. Rev. Lett. {\bf 103} 237001 (2009)

\bibitem{Kwon_2004} H.-J. Kwon, K. Sengupta, Victor M. Yakovenko, The European Physical Journal B \textbf{37}, 349(2004);
H.-J. Kwon, V. M. Yakovenko, and K. Sengupta, Low Temp. Phys. \textbf{30}, 613
(2004)




\bibitem{Tewari_PRL_2012}S. Tewari, J. D. Sau, Phys. Rev. Lett. 109,
150408 (2012)

\bibitem{Tewari_PRB_2012} S. Tewari, T.D. Stanescu,
J.D. Sau, S. Das Sarma, Phys. Rev. B 86, 024504 (2012)

\bibitem{Diez_2012} M. Diez, J. P. Dahlhaus, M. Wimmer, C. W. J. Beenakker, Phys. Rev. B \textbf{86}, 094501 (2012)







\end{thebibliography}
\end{document}